\documentclass{appolb}
\usepackage{epsfig}

\begin{document}
\title{LOOKING FOR LEPTONIC CP VIOLATION WITH NEUTRINOS%
\thanks{Slightly expanded written version of the talk presented at XXX Mazurian Lakes Conference on Physics, Piaski, Poland, September 2-9, 2007.}%
}
\author{Hisakazu MINAKATA
\address{Department of Physics, Tokyo Metropolitan University, 
Hachioji, Tokyo 192-0397, Japan}
}
\maketitle

\begin{abstract}

I discuss some theoretical aspects of how to observe leptonic 
CP violation. It is divided into two parts, one for CP violation due to 
Majorana, and the other more conventional leptonic 
Kobayashi-Maskawa (KM) phases. 
In the first part, I estimate the effect of Majorana phase to 
observable of neutrinoless double beta decay experiments by 
paying a careful attention to the definition of the atmospheric scale 
$\Delta m^2$. 
In the second part, I discuss Tokai-to-Kamioka-Korea two detector 
complex which receives neutrino superbeam from J-PARC as a 
concrete setting for discovering CP violation due to the KM phase, 
as well as resolving mass hierarchy and the $\theta_{23}$ 
octant degeneracy.  
A cautionary remark is also given on comparison between various 
projects aiming at exploring CP violation and the mass hierarchy. 

\end{abstract}

\section{Introduction}

On the occasion of 50 years anniversary of discovery of parity violation, 
this conference is focused on the problem of fundamental symmetries. 
Leptonic CP violation is an important and indispensable element 
of our understanding of not only neutrinos but also particle physics itself. 
It is because neutrino masses and the flavor mixing discovered by 
the atmospheric \cite{SKatm}, the solar \cite{solar}, and the reactor  
\cite{KamLAND} experiments constitute so far the unique evidence 
for physics beyond the Standard Model. 
Moreover, it may be the key to understand the baryon number 
asymmetry in the universe \cite{leptogenesis}.

CP violation in the lepton sector can be classified into 
the two categories;  
the one due to possible Majorana phase \cite{Mphase} and 
to the conventional Kobayashi-Maskawa (KM) phase \cite{KM} 
in the lepton flavor mixing matrix, the MNS matrix \cite{MNS}. 
While it is generally believed that CP violation of the latter type 
exists in nature, CP violation of the former type requires 
the existence of Majorana mass term. 
But, once neutrinoless (0$\nu$) double beta decay is observed, 
the Majorana CP violation must exist because of 
the general theorem \cite{generalTh}; presence of 
the 0$\nu$ double beta decay matrix elements implies the 
existence of the Majorana mass term irrespective of the origin 
of the 0$\nu$ double beta decay. 
Furthermore, there is a general argument \cite{yanagida} 
which states that under the assumption of Standard Model, 
neutrinos must be Majorana particles to explain 
the baryon asymmetry in the universe.

Despite that there is little doubt on the importance of uncovering leptonic 
CP violation, executing the task is extremely difficult. 
Therefore, the question of ``how to discover CP violation'' is 
worth to be discussed even more extensively than the level 
it has been done. 
In the following, I want to discuss some aspects of measuring 
CP violation due to the Majorana and the KM type phases 
under the hope that accumulating such discussions eventually 
leads to the feasible and promising experimental ideas.

\vspace{- 0.3cm}
\section{Measuring the Majorana phase}

Let us start with the discussion of the Majorana phase. 
Practically, 0$\nu$ double beta decay is the only known 
experimentally feasible way to identify the Majorana nature of 
neutrinos and measure the value of Majorana phase through its 
CP conserving effect \cite{kayser} in the rate. 
For recent reviews of 0$\nu$ double beta decay, see e.g. \cite{dbeta-review}.\footnote{
Apart from the possibility of observing Majorana CP violation 
the 0$\nu$ double beta decay is a rich source of informations. 
For example, if neutrinos are the Majorana particles and 
the degenerate mass spectrum is the case, it was shown that 
the small angle MSW solution is disfavored \cite{MYadn}. 
If the degenerate mass Majorana neutrinos exist, 
as claimed by Klapdor {\it et al.} \cite{klapdor},  
it might have been one of the first indications that the solar mixing 
angle is large. Of course, the claim in \cite{klapdor} has to be 
verified by the independent measurement. }

\subsection{Do 0$\nu$ double beta decay experiments distinguish 
the mass hierarchy?}

\begin{figure}[htb]
\begin{center}
\epsfig{file=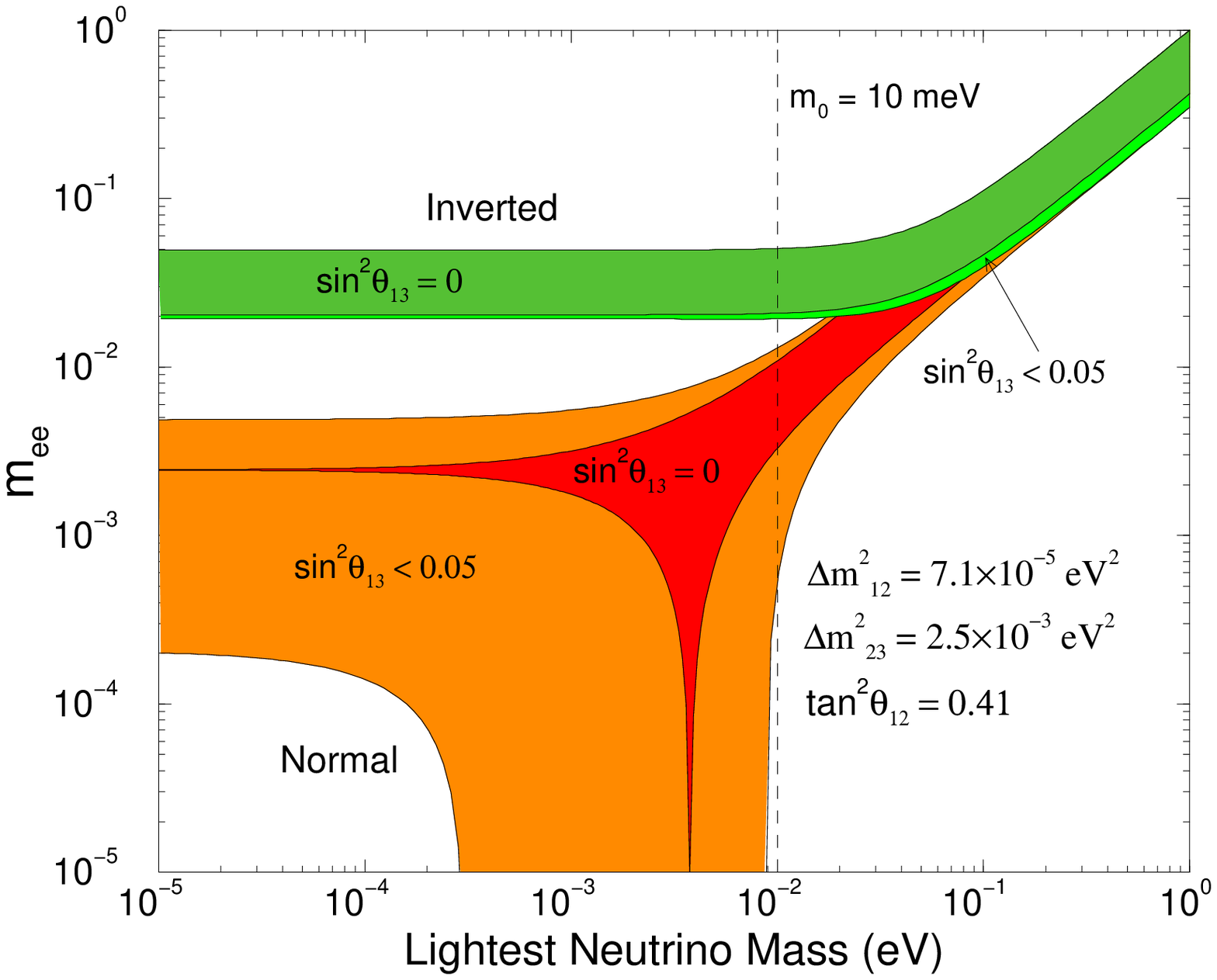,width=2.4 in}
\epsfig{file=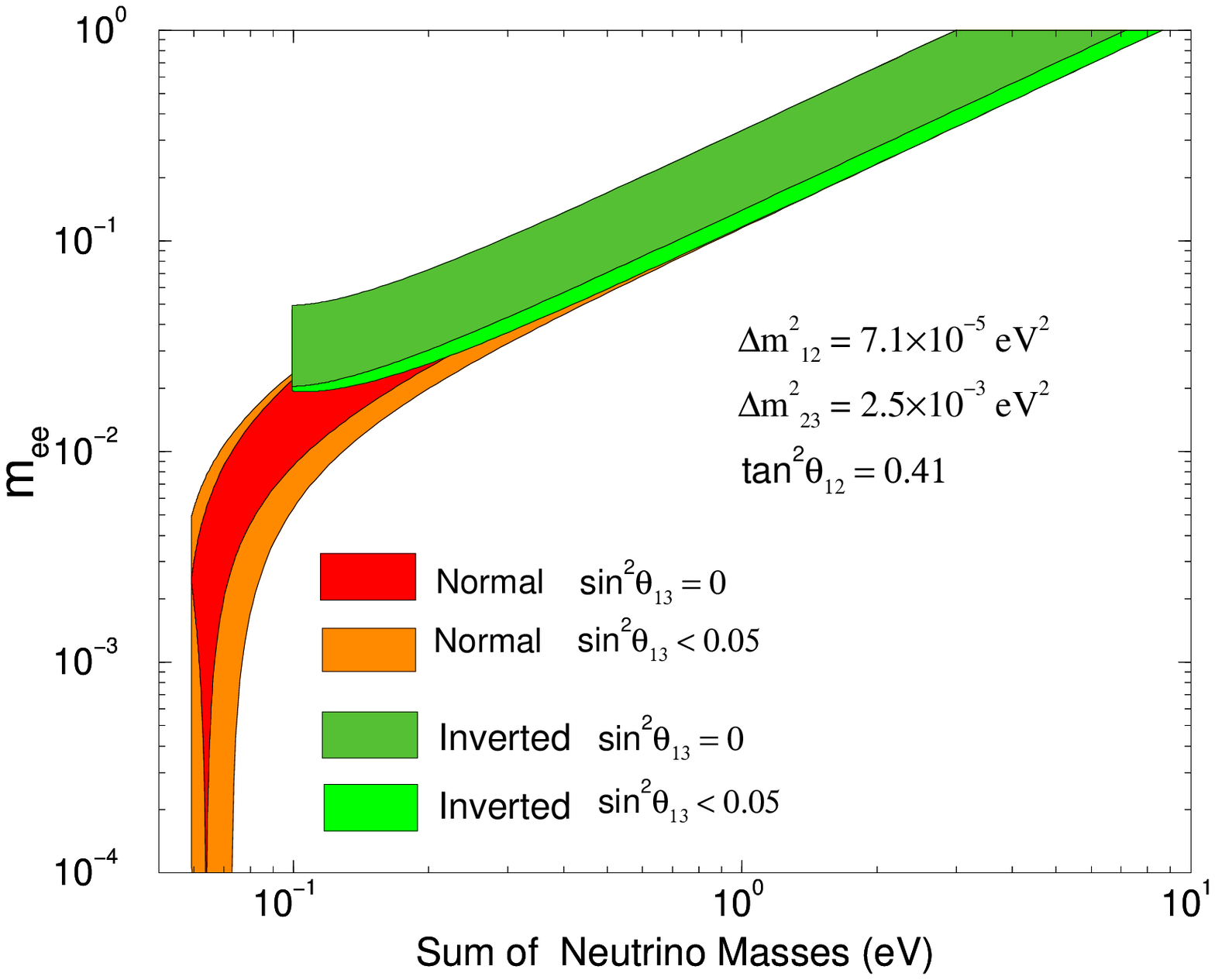,width=2.4 in}
\caption{The 0$\nu$ double beta decay observable 
$m_{ee} \equiv \langle m_{\beta \beta} \rangle$ in our notation, 
is plotted as a function of the lowest neutrino mass (left panel) 
and of sum of neutrino masses, 
$\Sigma \equiv \sum_{i=1}^{3} m_{i}$ (right panel). 
The figures are by courtesy of Hiroshi Nunokawa. 
}
\label{m0mee_bestfit.eps}
\vspace{-0.2cm}
\end{center}
\end{figure}

In Fig.~\ref{m0mee_bestfit.eps} plotted is the 0$\nu$ double beta 
decay observable $ \langle m \rangle_{\beta \beta} $ as a function of 
the lowest neutrino mass (left panel) 
and of sum of neutrino masses, 
$\Sigma \equiv \sum_{i=1}^{3} m_{i}$ (right panel). 
It appears that because of the clear distinction between the 
normal and the inverted mass hierarchies in the left figure 
the double beta decay experiments can discriminate between 
the hierarchies. However, the picture changes completely if 
one looks at the right panel of Fig.~\ref{m0mee_bestfit.eps}. 
Therefore, as far as $\Sigma$ is the only additional observable, 
it can be done only in a tiny region, unfortunately.

Therefore, I would like to take the attitude that the neutrino mass 
hierarchy will be determined by future accelerator LBL experiments, 
and consider below the implication of possible future observation of 
0$\nu$ double beta decay events in the context of Majorana CP violation. 
(See Sec.~\ref{T2KK} for an example of such LBL projects.)

\subsection{Analytic estimate of the effect of Majorana phase}

With use of the standard notation of the MNS matrix \cite{PDG}, 
the observable in neutrinoless double beta decay 
experiments can be expressed as 
\begin{eqnarray}
\langle m \rangle_{\beta \beta} &=& 
\left\vert m_1 c_{12}^2c_{13}^2 e^{ i \phi_{1} }
+ m_2 s_{12}^2c_{13}^2 e^{+i \alpha}
+ m_3 s_{13}^2 e^{ - i \alpha }
\right\vert, 
\nonumber \\
&=& 
\left\vert m_1 c_{12}^2c_{13}^2 e^{ + i \beta}
+ m_2 s_{12}^2c_{13}^2 e^{- i \beta}
+ m_3 s_{13}^2 e^{i \phi_{3} }
\right\vert, 
\label{mee}
\end{eqnarray}
where $m_i$ (i=1, 2, 3) denote neutrino mass eigenvalues. 
The first and the second lines of (\ref{mee}) are for the 
normal ($\Delta m^2_{32}>0$) and the 
inverted ($\Delta m^2_{32}<0$) mass hierarchies, respectively, 
and the Majorana phases $\alpha$, $\beta$ etc. are parametrized 
in (\ref{mee}) in a convenient way for our later discussions \cite {MYadn}.
We use the convention that $m_3$ is the largest (smallest) mass 
in the normal (inverted) mass hierarchy.

To express the mass eigenvalues in terms of observable 
$\Delta m^2$'s we pay careful attention to the fact that 
neither $|\Delta m^2_{32}|$ nor $|\Delta m^2_{31}|$ 
(with definition of $\Delta m^2_{ij} \equiv m^2_{i} - m^2_{j}$) 
is the observable quantity. In $\nu_{\mu}$ disappearance 
measurement it is \cite{NPZ}\footnote{
If we use $\nu_{e}$ disappearance measurement it is given by 
$\Delta m^2_{ee} = c^2_{12} \vert \Delta m^2_{31}  \vert + 
s^2_{12} \vert \Delta m^2_{32} \vert $ \cite{NPZ}. 
These expressions are shown to be of key importance in estimating 
sensitivities to mass hierarchy resolution by the disappearance 
methods \cite{MNPZ}.
}
%
\begin{eqnarray}
\Delta m^2_{\mu \mu } = 
s^2_{12} | \Delta m^2_{31} | + c^2_{12} |\Delta m^2_{32} | + 
\cos \delta \sin 2 \theta_{12} \tan \theta_{23} s_{13} \Delta m^2_{21}. 
\label{Dm2mumu}
\end{eqnarray}
Solving (\ref{Dm2mumu}) we obtain for the normal mass hierarchy 
($\Delta m^2_{32} > 0$, $m_1 = m_{\it l}$);
\begin{eqnarray}
m^2_{3} &=& \Delta m^2_{\mu \mu } 
\left[ 1 + 
\epsilon \left\{
c^2_{12} - \cos \delta \sin 2 \theta_{12} \tan \theta_{23} s_{13} 
\right\} + \kappa 
\right], 
\nonumber \\
m^2_{2} &=&  \Delta m^2_{21} 
\left[ 1 + \frac{ \kappa } { \epsilon } \right]
\label{m2normal}
\end{eqnarray}
and for the inverted mass hierarchy 
($\Delta m^2_{32} < 0$, $m_3 = m_{\it l}$);
\begin{eqnarray}
m^2_{2} &=& \Delta m^2_{\mu \mu } 
\left[ 1 + 
\epsilon \left\{
s^2_{12} - \cos \delta \sin 2 \theta_{12} \tan \theta_{23} s_{13} 
\right\} + \kappa 
\right], 
\nonumber \\
m^2_{1} &=& \Delta m^2_{\mu \mu } 
\left[ 1 - 
\epsilon \left\{
c^2_{12} + \cos \delta \sin 2 \theta_{12} \tan \theta_{23} s_{13} 
\right\} + \kappa 
\right], 
\label{m2inverted}
\end{eqnarray}
where $\epsilon \equiv \Delta m^2_{21} / \Delta m^2_{\mu \mu } \simeq 0.032$ 
and $\kappa \equiv m^2_{\ell} / \Delta m^2_{\mu \mu } $.

\subsection{Observable under the approximation of ignoring lowest $\nu$ mass}

For simplicity, we restrict our discussion to the case $m_{\ell}$ can be 
ignored compared to other two mass eigenvalues. 
Then, we obtain the expression of 
$ \langle m \rangle _{\beta \beta}^2$ for the 
normal mass hierarchy as 
\begin{eqnarray}
\frac{ \langle m \rangle_{\beta \beta}^2 }{ \Delta m^2_{21} } &=& 
s^4_{12} c^4_{13} + 
\frac{ s^4_{13} }{ \epsilon } + 
 2 \sqrt{ \frac{ s^4_{13} }{ \epsilon } }  s^2_{12} c^2_{13} \cos 2\alpha
+ \sqrt{  \epsilon } s^2_{13}   c^2_{12} s^2_{12}  \cos 2\alpha, 
\label{mee-normal}
\end{eqnarray}
where we have ignored the terms of order $\epsilon^2$, 
$s^4_{13}$ which are not enhanced by inverse power of 
$\epsilon$, and $\sqrt{ \epsilon } s^3_{13}$. 
Notice that $\langle m \rangle_{\beta \beta}^2$ is naturally of 
order $\sim \Delta m^2_{21} $ in the normal hierarchy. 
In the inverted hierarchy,  it is of the order of $\Delta m^2_{atm} $ 
and takes the form 
\begin{eqnarray}
\frac{ \langle m \rangle_{\beta \beta}^2 }{ \Delta m^2_{\mu \mu }   } &=&  
c^4_{13}
\left[ 1 - \sin^2 2\theta_{12} \sin^2 \beta - 
\epsilon \left\{ 1 - \frac{1}{2} \sin^2 2\theta_{12} \sin^2 \beta \right\} 
\right.
\nonumber \\
&&\hspace*{12mm} 
\left.
- \epsilon \cos \delta \sin 2 \theta_{12} \tan \theta_{23} s_{13} 
\left\{ 1 -  \sin^2 2\theta_{12} \sin^2 \beta \right\} 
\right], 
\label{mee-inverted}
\end{eqnarray}

Here, we notice that the difference between our results in 
(\ref{mee-normal}) and (\ref{mee-inverted}) and the ones which 
would be obtained if we use the naive definition 
$\Delta m^2_{atm} = \Delta m^2_{32} $ is very small. 
In the normal hierarchy case, it is of the order of the last term 
in (\ref{mee-normal}) (coefficient doubled) which is 
of order $\sqrt{ \epsilon } s^2_{13} \leq 5 \times 10^{-3}$. 
In the inverted hierarchy case, the difference is in order 
$\epsilon \simeq 0.03$ terms; 
The second line in (\ref{mee-inverted}) is of course missing 
and unity in the curly bracket in the first line is replaced by $s^2_{12}$. 
Therefore, the careful definition of the atmospheric $\Delta m^2$ 
\cite{NPZ} does not appear to produce detectable difference 
in the 0$\nu$ double beta decay observable. 
This feature seems to be generic and is true without approximation 
of ignoring the lowest neutrino mass.

We estimate how large is the effect of the Majorana phase in 
(\ref{mee-normal}) and (\ref{mee-inverted}). 
To make a fair comparison between the normal and the inverted 
hierarchy cases we compute the ratio of the coefficient 
of $\cos 2\alpha$ (or $\cos 2\beta$) to the phase independent piece, 
$B/A$ in $\langle m \rangle_{\beta \beta}^2 = A + B \cos 2 \alpha$. 
(Note that the experimental observable is the square of 
$\langle m \rangle_{\beta \beta}$.) 
The results of the ratios are about 
$0.72\times(s^2_{13} / 0.025) $  and 0.79 independent of $s_{13}$ 
in the normal and the inverted mass hierarchies, respectively. 
Therefore, the effect of the Majorana phase is large in both hierarchies 
(for the normal hierarchy if $s^2_{13}$ is large) and 
should be observed if the experiments are accurate enough and 
the uncertainty of the nuclear matrix elements can be solved. 
It is the challenge to the next generation double beta decay 
experiments, whose partial list is in \cite{dbeta-next}, to reach 
the sensitivity of this level to observe the Majorana phase. 
A promising idea for resolving the problem 
(or at least much improving the situation)  
of uncertainty of the nuclear matrix elements is proposed \cite{rodin}.

What about the uncertainties which may be caused by other mixing 
parameters involved in (\ref{mee-normal}) and (\ref{mee-inverted}). 
The error of $s^2_{12}$ can be controlled down to about 
a few \% if low energy $pp$ neutrino is accurately 
measured \cite{bahcall-pena}, and/or to $\simeq$2\% 
if a dedicated reactor \cite{sado} or  the M\"ossbauer-type 
measurement \cite{mina-uchi_NJP} is executed. 
The error for $\sin^2 2\theta_{12}$ is even smaller by a factor of $\sim$2. 
Therefore, there is little additional uncertainty in the inverted hierarchy case. 
In the normal hierarchy case the major uncertainty would come from 
the error of $s^2_{13}$ measurement on which we do not yet have 
definitive perspective.

Though my analysis in this paper is rather qualitative I hope 
it illuminates the main point of the more quantitative analysis. 
Examples of recent analysis of 0$\nu$ double beta decay, 
in particular on the possibility of observing the Majorana phase 
see e.g.  \cite{dbeta-phase}.

\vspace{- 0.3cm}
\section{Measuring the leptonic Kobayashi-Maskawa phase}

\subsection{General comments}

Probably the most popular way of measuring CP violation 
of the KM type is the long-baseline (LBL) accelerator neutrino 
experiments. 
Since the appearance oscillation probabilities of 
$\nu_{\mu} \rightarrow \nu_{e}$ and the one with their anti-neutrinos 
have different dependence on CP phase $\delta$, 
$P(\nu_{\mu} \rightarrow \nu_{e}) - 
P(\bar{\nu}_{\mu} \rightarrow \bar{\nu}_{e}) = 16 
c_{12} s_{12} c_{23} s_{23} c_{13}^2 s_{13}  
\Pi_{i, j= {\rm cyclic} } \sin \left( \frac{\Delta m_{ij}^2 L}{4E} \right)$ 
in vacuum, 
one can in principle detect the effect of CP violation by comparing 
$\nu_{e}$ ($\bar{\nu}_{e}$) yields in $\nu_{\mu}$ ($\bar{\nu}_{\mu}$) 
beam exposure.

Unfortunately, it is not the end of the story. 
The earth matter effect inevitably comes in as a contamination 
in CP phase measurement, because 
the earth matter is CP asymmetric. 
The interplay between the genuine CP phase effect and 
the matter effect is extensively discussed in the literature.  
A few early ones are in \cite{AKS,MNprd98,golden}. 
One way of dealing with the issue of matter effect 
contamination is to carry out the experiments in a near vacuum setting, 
low energy conventional superbeam to search for CP violation 
\cite{lowECP}. 
Concrete examples of such setting include the ones described in 
\cite{MEMPHYS,T2K}. 
Another way to deal with the problem is to perform {\em in situ} 
measurement of the matter effect in the experiments, 
as emphasized in \cite{Nutele07} and illustrated for neutrino 
factory in \cite{mina-uchi,gandhi-winter}. 
Effects of errors in the matter density on the CP sensitivity in 
neutrino factory is discussed e.g., in \cite {huber_optim}.

Nonetheless, it is important to recognize that the matter contamination in 
CP violation measurement is {\em not} entirely a bad news. 
Namely, one can resolve the neutrino mass hierarchy by 
utilizing the interference between the vacuum and the matter effects. 
I think that these discussions make it clear that we must invent 
a consistent experimental framework in which one can achieve 
simultaneous determination of CP phase $\delta$ and the 
neutrino mass hierarchy. 
Furthermore, it became the {\em necessity} when the problem of 
parameter degeneracy, the ambiguity due to multiple solutions of 
lepton mixing parameters, was uncovered \cite{intrinsic,MNjhep01,octant}. 
For its various aspects, see \cite{BMW,KMN,MNP2}. 
Unless we can formulate the way of how to resolve it and give reliable 
estimation of the sensitivity it would be difficult to convince 
people of the value of such an inevitably expensive project.

\subsection{T2KK; Tokai-to-Kamioka-Korea two detector complex}
\label{T2KK}

I want to describe our proposal which we believe to possess the 
required properties as described above. It is now called 
T2KK, acronym of the Tokai-to-Kamioka-Korea identical two 
detector complex \cite{T2KK1st,T2KK2nd}. 
See \cite{T2KKweb} for more global overview of the project, 
in particular the latest status as well as atmosphere in the initial stage.

\begin{figure}[htb]
\begin{center}
\epsfig{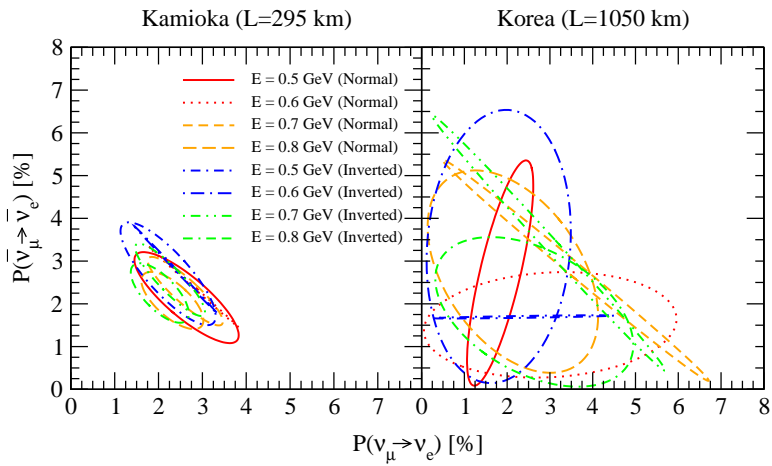}
\caption{Energy dependences of the oscillation probabilities 
for $\sin^2 2\theta_{13}=0.05$ are represented by plotting ellipses 
(which results as $\delta$ is varied from 0 to $2\pi$) 
in bi-probability space for various neutrino energies from 0.5 to 0.8 GeV.
The left and the right panels are for detectors in Kamioka and in 
Korea, respectively. 
The ellipses in upper 4 symbols (warm colors) indicate the ones of 
normal mass hierarchy ($\Delta m^2_{31} >0$)
and the one of lower 4 symbols (cold colors) the ones of 
inverted mass hierarchy ($\Delta m^2_{31} <0$). 
The figure is taken from \cite{T2KK1st}.
}
\label{ellipse-kam-korea}
\vspace{-0.2cm}
\end{center}
\end{figure}

The basic idea of T2KK setting is the comparison between the 
yields at the two detectors \cite{MNplb97}, 
one in Kamioka (295 km) and the 
other in Korea (1050 km) whose former (latter) location is 
near the first (second) oscillation maximum. 
As indicated in Fig.~\ref{ellipse-kam-korea} in the form of 
$P(\nu_{\mu} \rightarrow \nu_{e})$-$P(\bar{\nu}_{\mu} \rightarrow \bar{\nu}_{e})$ bi-probability plot \cite{MNjhep01} the behavior of the 
appearance oscillation probabilities in Kamioka and Korea are 
vastly different. 
It will allow us to resolve the intrinsic \cite{intrinsic} and the 
sign-$\Delta m^2_{31}$  \cite{MNjhep01} degeneracies to 
determine CP phase $\delta$ and the mass hierarchy.

\begin{figure}[htb]
\begin{center}
\epsfig{file=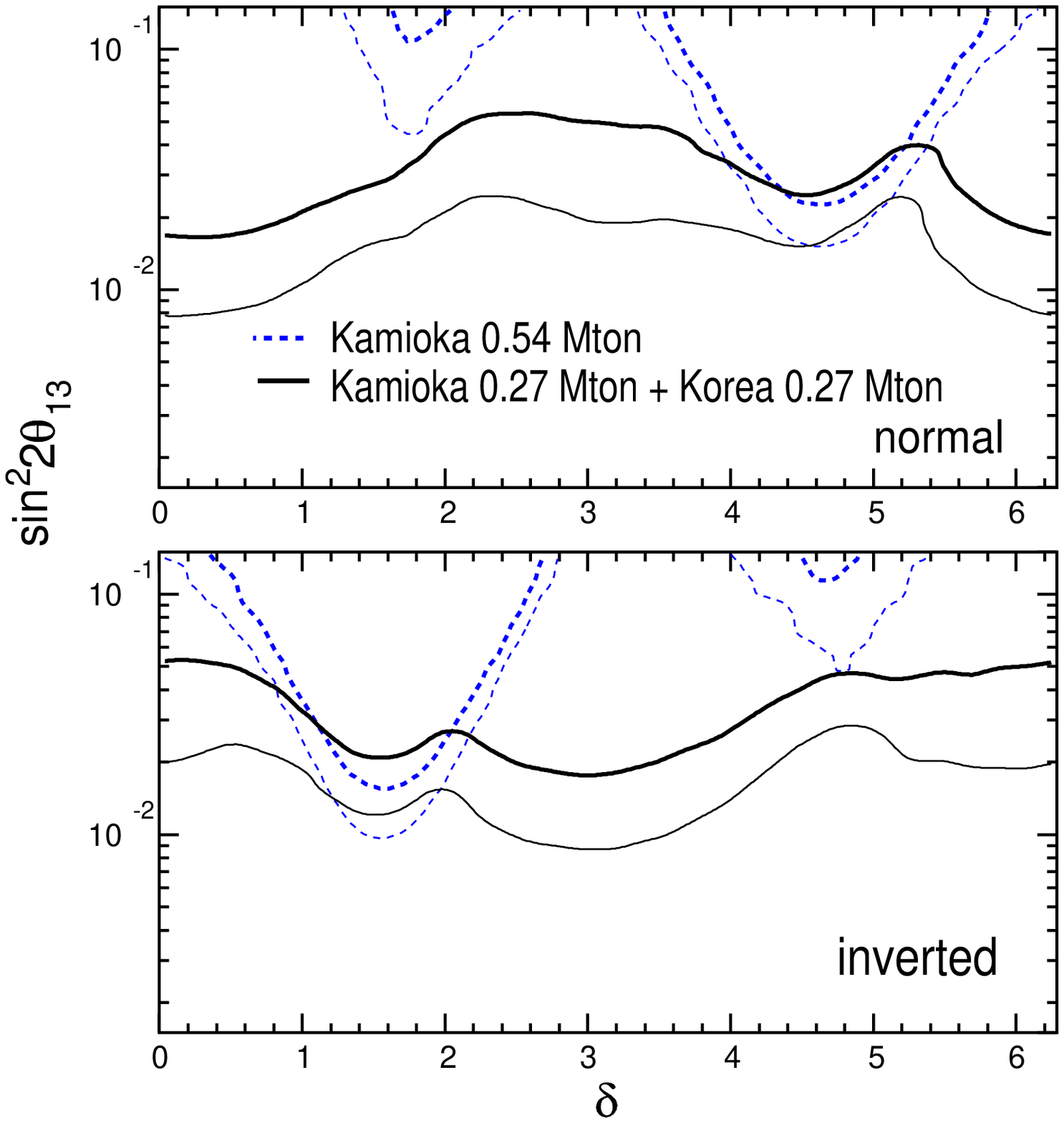,height=2.6 in}
\epsfig{file=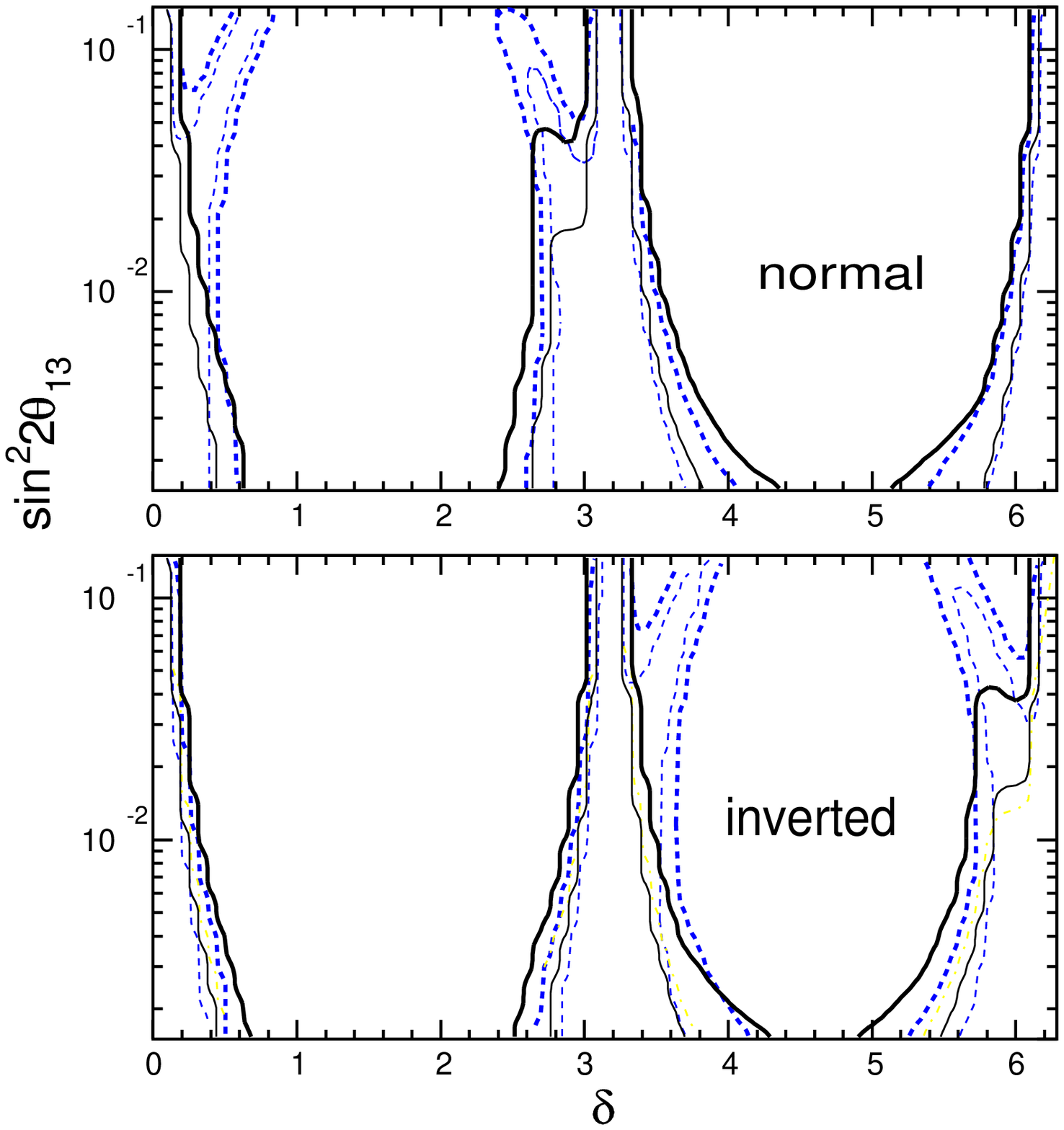,height=2.6 in}
\caption{Presented are the sensitivities to the mass hierarchy 
determination (left panel) and CP violation (right panel) at 
2$\sigma$(thin lines) and 3$\sigma$ (thick lines) CL.  
The standard deviation is defined with 1 degree of freedom (DOF).
The black solid lines are for the T2KK setting while 
the blue dotted lines are for the T2K II setting. 
$\theta_{23}$ is taken to be maximal. 
The top and bottom panels are for the normal and 
the inverted mass hierarchies, respectively.
The results is from \cite{T2KK1st}. 
}
\vspace{-0.2cm}
\label{sensitivity_mass-CP}
\end{center}
\end{figure}

In Fig.~\ref{sensitivity_mass-CP} presented are the results of the 
sensitivities to mass hierarchy resolution and detection of CP 
violation obtained in \cite{T2KK1st}. 
As indicated in the figure, the T2KK setting has potential of discovering 
KM type leptonic CP violation and at the same time resolve the neutrino 
mass hierarchy if $\theta_{13}$ is in the region of sensitivities 
possessed by the next generation reactor \cite{3projects} and the 
accelerator LBL experiments \cite{T2K,NOVA}.

I also note that T2KK has potential of lifting the $\theta_{23}$ octant 
degeneracy \cite{T2KK2nd} by detecting the solar scale oscillation 
effect, as proposed for atmospheric neutrino observation \cite{atm-method}. 
The sensitivities to the octant ambiguity resolution is better (worse) 
compared to the best thinkable sensitivity achievable by the 
reactor-accelerator combined method \cite{MSYIS} in a region 
$\sin^2 2 \theta_{13}$ smaller (greater) than 0.05-0.06.  
These values are based on the sensitivity 
estimated in \cite{T2KK2nd} and \cite{hiraide}. 
To sum up, T2KK will have an {\em in situ} potential of resolving 
the total eight-fold parameter degeneracy, thereby fulfilling 
the required qualification as a candidate for future LBL neutrino 
oscillation experiments for determining lepton mixing parameters.

Since most of the future LBL projects are equipped with large detectors 
they automatically possess the potential of resolving the 
$\theta_{23}$ octant degeneracy by atmospheric neutrino observation. 
The point is, however, that by having an {\em in situ} potential 
of resolving the degeneracy T2KK can use such the additional 
capability as a consistency check of the results, guaranteeing 
the desirable ``redundancy''. 
Since the systematic errors involved are quite 
different in both methods I believe that such redundancy must be 
retained to make the measurement robust ones. 
The similar statement may apply to the mass hierarchy resolution.

\subsection{Remarks on comparison between the projects}

Some people tries to compare the sensitivities to the mass 
hierarchy resolution and CP violation possessed by various 
projects proposed \cite{huber-NNN07}. 
Though important I would like to make a cautionary remark 
on such comparison between projects which use water 
Cherenkov detectors. 
It is known that the issue of background rejection at high energies 
becomes highly nontrivial in the detector. 
Therefore, if one wants to compare two settings which does 
and does not require the special care for background rejection 
at high energies, this problem has to be settled first in a convincing way. 

\begin{figure}[htb]
\begin{center}
\epsfig{file=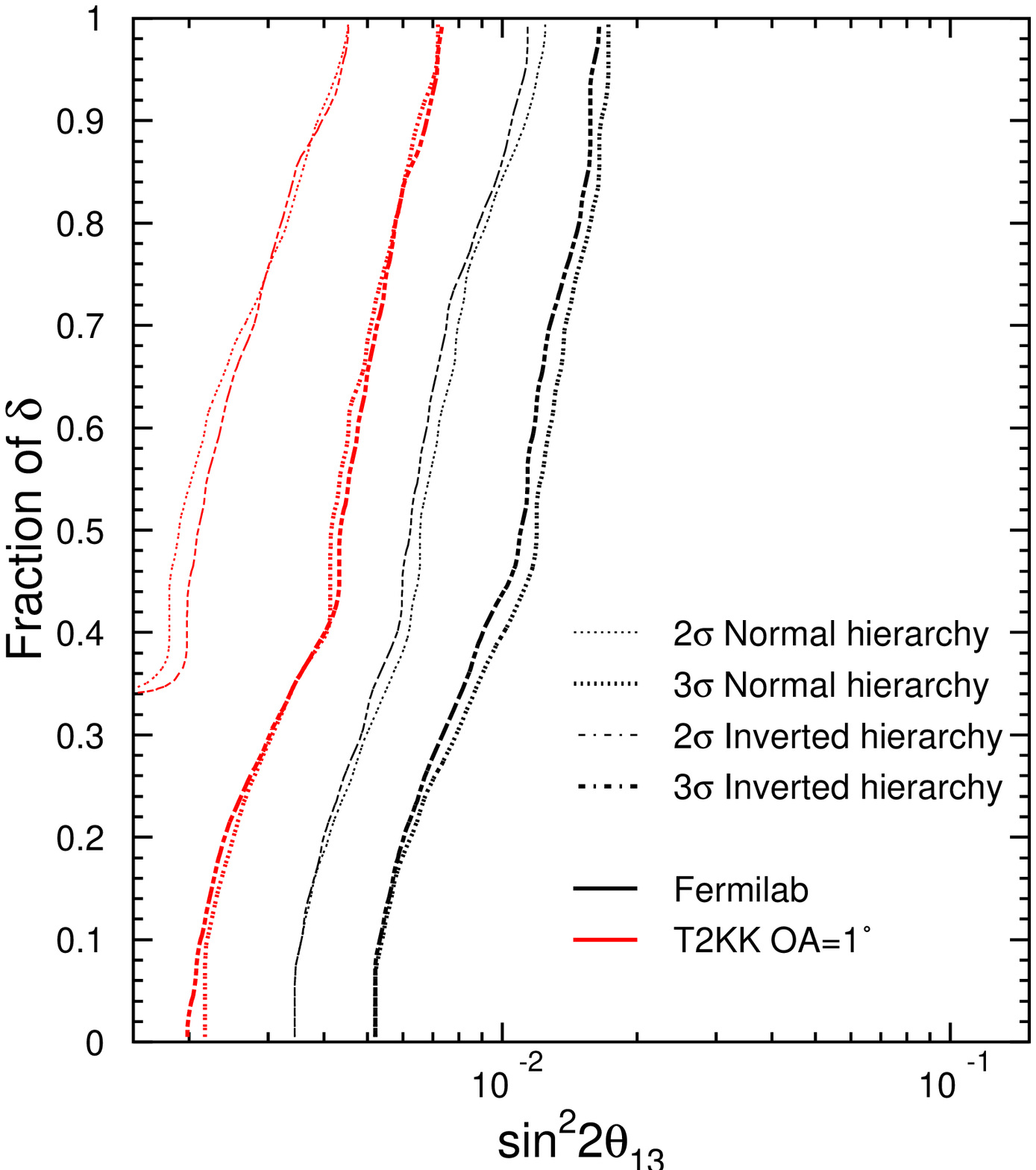,height=2.4 in}
\epsfig{file=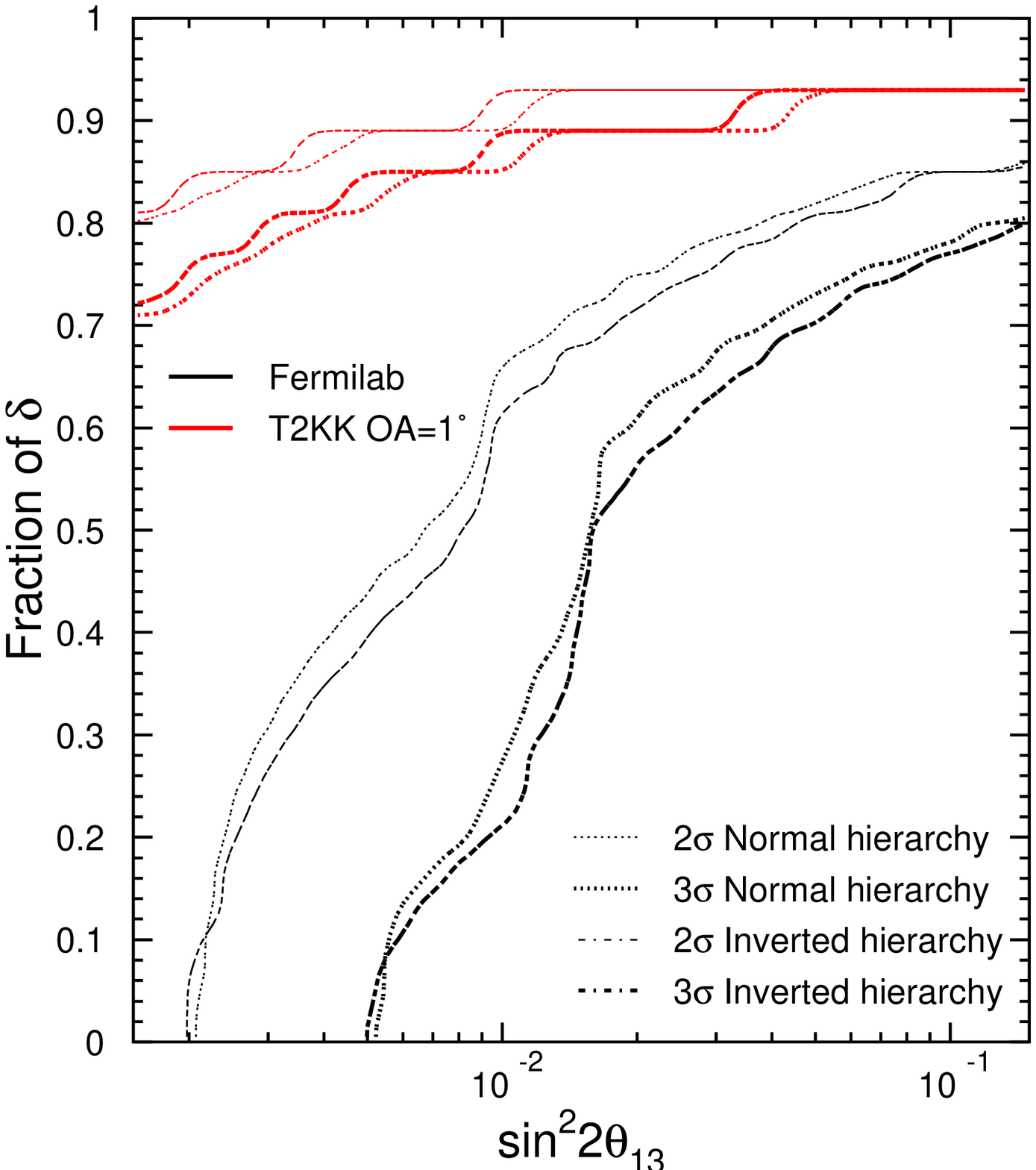,height=2.4 in}
\caption{Fractional coverage of CP phase $\delta$ in which 
there are sensitivities to the mass hierarchy 
determination (left panel) and CP violation (right panel) 
by Fermilab-Homestake VLBL project (black curves) and 
T2KK with Korean detector at 1 degree off-axis angle (red curves) 
\cite{dufour-T2KK3rd}. 
The thick (thin) lines are at 3$\sigma$ (2$\sigma$) CL. 
The standard deviation is defined with 1 DOF.
$\theta_{23}$ is taken to be maximal. 
The dotted and the dash-dotted lines are for the normal and 
the inverted mass hierarchies, respectively. 
The figure is by courtesy of Fanny Dufour.
}
\label{sensitivity-comparison}
\end{center}
\end{figure}

As an example I show in Fig.~\ref{sensitivity-comparison} 
the sensitivity estimate done by Dufour \cite{dufour-T2KK3rd} 
for the VLBL project with the 
Ferimilab-Homestake baseline (1300 km) 
If one compares Fig.~\ref{sensitivity-comparison} 
to Fig.~11 in \cite{barger}, 
one notices that the sensitivity reach for the mass hierarchy 
obtained by Dufour is worse than that in \cite{barger} 
despite usage of more aggressive setting of 
beam power of 2 MW for 5 years running for both neutrinos 
and antineutrinos. For more details, see \cite{dufour-T2KK3rd}. 
It should be noticed that while Fig.~\ref{sensitivity-comparison} 
contains a comparison between the discovery potentials of the 
two projects, T2KK and the VLBL project, the settings of the both 
experiments (beam power etc.) are rather different.

\section{Concluding remarks}

In this talk, I tried to describe some aspects of the problem of how to 
detect CP violation due to both the Majorana and the KM phases 
in the lepton mixing. 
Though they are extremely difficult to carry out, the implications of 
the detection are so great that it is worth to continue to think about them. 
%
I thank the organizers for invitation to the conference in such 
a scenic place, Mazurian Lakes, which gave me the chance 
of revisiting the issue of Majorana phase in the 0$\nu$ double beta decay.

\section*{Acknowledgments}

I thank Stephen Parke for his useful comments, and 
to Fanny Dufour and Hiroshi Nunokawa for their 
kind permission of using the VLBL sensitivity and the double 
beta decay figures. 
This work was supported in part by KAKENHI, Grant-in-Aid for 
Scientific Research, No 19340062, Japan Society for the 
Promotion of Science.

\end{document}